\documentclass[guide]{rmaa}
\usepackage{booktabs,graphics,graphicx,keyval,natbib,paralist,psfrag,trig,supertabular}
\title{ON THE MAGNETIC FIELD EVOLUTION IN SHELL-LIKE SUPERNOVA
    REMNANTS}
\author{B. Vukoti{\' c}\altaffilmark{1}, B. Arbutina\altaffilmark{1} and D. Uro{\v s}evi{\'c}\altaffilmark{2}}

\altaffiltext{1}{Astronomical Observatory, Belgrade, Serbia }

\altaffiltext{2}{Department of Astronomy, Faculty of Mathematics,
University of Belgrade, Serbia}

 \fulladdresses{
\item B. Vukoti{\' c} and B. Arbutina: Astronomical Observatory, Volgina 7, 11160 Belgrade 74, Serbia
(bvukotic@aob.bg.ac.yu, barbutina@aob.bg.ac.yu)

\item D. Uro{\v s}evi{\'c}: Department of Astronomy, Faculty of Mathematics,
University of Belgrade, Studentski Trg 16, 11000 Belgrade, Serbia
(dejanu@poincare.matf.bg.ac.yu)

}

\shortauthor{VUKOTI{\' C} ET AL.} \shorttitle{ON THE MAGNETIC
FIELD EVOLUTION IN SUPERNOVA
    REMNANTS}

\ReceivedDate{2006 January 26} \AcceptedDate{2006 October 6}
\SetYear{2006} \resumen{ \small En este papel aplicamos y
discutimos un m\'etodo para la determinaci\'on� de la evoluci\'on� del
campo magn\'etico en los remanente de la supernova (SNRs) de la
luminosidad de radio en la frecuencia $\nu$ dada a la correlaci\'on�
del di\'ametro ($L_\mathrm{\nu}-D$). Asumimos que $H$ se desarrolla
como $H\propto D^{-\delta}$, donde $D$ est\'a�el di\'ametro del
remanente. El valor $\delta\approx1.2$ se obtiene bajo asunci\'on�
del equipartition de las ecuaciones para el c\'alculo revisado del
equipartition (REC) y usar la muestra de los datos de la galaxia
pr\'oxima M82 del starburst. Intentamos investigar si o no los remanente de la supernova en M82
est� en el estado del equipartition.
Inconsistence entre el valor obtenido te\'oricamente para las condiciones y la
extensi\'on� adib\'atica $(\delta=1.5)$ del equipartition y el valor
emp�ico obtenido adjunto, se puede explicar principalmente por la
influencia de los efectos de la selecci\'on� de la sensibilidad que
tienden para aplanar la cuesta de las relaciones del $L_\nu-D$
para las muestras extragalactic. }

\abstract{\small In this paper we apply and discuss a method for
 the determination of the magnetic field ($H$) evolution in
supernova remnants (SNRs) from radio luminosity at given
frequency $\nu$ to diameter ($L_\mathrm{\nu}-D$) correlation. We
assumed that $H$ evolves as $H\propto D^{-\delta}$, where $D$ is
the diameter of the remnant.
 Value $\delta\approx1.2$ is
obtained under the equipartition assumption from the equations for
revised equipartition calculation (REC) and  by using the
data sample from the nearby starburst galaxy M82.  We try to investigate whether or
not SNRs in M82 are in the equipartition state. This is done by
comparison of our empirically obtained $\delta$ with the
theoretical value  expected for the equipartition conditions.
Inconsistence between the value obtained for equipartition
conditions and adiabatic expansion $(\delta=1.5)$ and the value
empirically obtained  herein, can be explained mainly by the
influence of sensitivity selection effects which tend to flatten
the slope of the $L_\nu-D$ relations for extragalactic samples.}


\begin{document}
\maketitle

\section{Introduction}

Supernova remnants (SNRs) stand an important factor in the process
of cosmic ray acceleration and  matter circulation.  Albeit
very important,  these processes are still not fully
understood. Various theories were  suggested during the last
few decades {with a view to understanding} the SNR properties.
There is a general belief that  the evolution of an SNR is
strongly influenced by the properties of the local interstellar
medium (ISM) in which SNR evolves.  As SNRs are the luminous
synchrotron emitters in radio domain of the electromagnetic
spectrum, the magnetic field
 inside  them and energetic spectrum of relativistic
particles can be  determined. Here, we will mainly  focus
on the magnetic field properties such as field strength and
evolution. The most commonly used empirical relation in studies of
SNR evolution properties is radio surface brightness to diameter
($\Sigma-D$) relation. This is because the only statistically
reliable data samples of SNRs are found in radio domain.  In
order to study SNR evolution issues from  a slightly
different perspective, in this paper we apply a method that
transforms $\Sigma-D$ into magnetic field to diameter ($H-D$)
relation. This way, we can  discuss SNR evolutionary
properties by comparing theories on $H$ with empirically extracted
$H-D$ relation.  Statistical i.e. empirical study of $H$
evolution  nevertheless requires reliable data samples 
of SNRs in different types of interstellar medium.

There  is a number of ways to estimate $H$ in SNRs.
Unfortunately,  few are reliable and  even they are
available only for a few well studied SNRs. The estimates are made
by measuring rotation measures or  spectral line splitting. The
estimates can also be extracted from radiation fluxes from
different parts of the electromagnetic spectrum such as radio,
X-rays or $\gamma$-rays. However, there is another problem in
performing   a statistical study of $H$  based on
these estimates. The data samples of SNRs are ballast by severe
selection effects through the  entire electromagnetic
spectrum. SNRs are mainly identified in the radio domain. Unlike
optical, X-rays or $\gamma$-rays, radio waves are less influenced
by absorption and scattering in the interstellar medium. Also,
radio interferometers have the best resolution among  all the
other observational devices, which also helps in  the
detection of remnants. Large and reliable data samples are of
crucial importance  for a good and well-founded  statistical
study  of empirical $H-D$ relation. Today, this condition is
partially fulfilled only  by data samples in radio domain.
The empirical studies of SNR properties are  also severely
influenced by the selection effects.   It remains to be hoped
that the  observational instruments and techniques in the future
will help us overcome this problem.

The main purpose of this paper is to apply and discuss a method
for  the determination of $H-D$ slope from radio luminosity
at  given frequency {$\nu$} to diameter ($L_\mathrm{\nu}-D$)
correlation ($\Sigma_\mathrm{\nu}=L_\mathrm{\nu}/(D^2\pi^2)$), in
SNR samples that show existence of such  a correlation. The
method is based on the energy equipartition assumption between
magnetic field and relativistic particles. It uses  equations for
revised equipartition calculation (REC). The equipartition
calculation is  the most commonly used manner of obtaining
$H$ estimates in valid radio SNR samples. The obtained $H-D$ slope
from the only reliable data sample of M82 SNRs is then compared
with the slope arising from the theoretical models of SNR
evolution.  We can then argue whether or not M82 SNRs are in
the equipartition state,  and thereby give a contribution to
the general evolutionary studies of SNRs. We also try to
give estimates of the magnetic field strengths, particularly for
SNRs in M82.  In addition, we discuss the accuracy of magnetic
field strength obtained under REC. This is done by  comparing
the values for $H$, obtained  herein, with the more reliable
ones available in literature (found for few SNRs from Large
Magellanic Cloud and our Galaxy).  It is noteworthy that an
SNR luminosity is mainly determined by the density of environments
in which SNR evolves. This is  an important issue for 
the discussion of  the influence of equipartition arguments
on $H$.

 This paper is organized as follows:
 Section 2 presents explanations of
 the required topics which are too broad to be mentioned later in the text.
 In Section 3 we describe and analyze the method and REC with its assumptions.
 Section  4  features a discussion on  the obtained results for $H-D$ slope and magnetic field strength.
 There, we  consider whether M82 SNRs are in the equipartition state  or not.
 Finally, the conclusions of this work are given in Section 5.

\section{The $H-D$ dependence}

\subsection{History of $H-D$ Relation}

We assume that  $H-D$ relation can  be written in the form:
\begin{equation}
H \propto D^{-\delta}.
\end{equation}
Historically, this form of the magnetic field evolution is used
 in all theoretical models that explain the synchrotron
emission from SNRs.

Shklovsky (1960) was  the first  to theoretically describe
the synchrotron emission from  a spherically expanding
nebula. He assumed that magnetic field structure remains unchanged
during the expansion. Consequently, magnetic field flux  is
constant and $H \propto D^{-2}$, where $D$ is the diameter of the
remnant. Lequeux (1962) applied Shklovsky's theory to model shell
type remnants, which  led to $H \propto (l \times D)^{-1}$,
where $l\propto D$ represents the thickness of the shell. Poveda
\& Woltjer (1968) and Kesteven (1968) also gave their contribution
to the general model of shell type remnant. They assumed that $H$
is gained with the compression of the interstellar  medium
 magnetic field (leading to $H=const$) and that shell
thickness remains constant during the expansion ( which leads
to $H \propto D^{-1}$). Theoretical interpretation of SNR
synchrotron emission by Duric \& Seaquist (1986) used the magnetic
field model with $H \propto D^{-\delta}$, based on the work of
Gull (1973) and Fedorenko (1983).  According to the results
of Gull, magnetic field is compressed and amplified in the
convection zone, to finally gain enough strength to power the
bright synchrotron emission. Fedorenko stated that $1.5 \le \delta
\le 2.0$. Tagieva (2002) obtained $H\propto D^{-0.8}$, by using the $\Sigma\propto D^{-2.38}$ relation (Case \& Bhattacharya 1998). However, this result should be taken with great reserve because the $\Sigma-D$ relation from the work of Case \& Bhattacharya is balast by severe selection effects (Uro{\v s}evi{\' c} et al. 2005). Also, Tagieva did not take into account the influence of the density of environments in which the SNRs evolve. An interesting discussion about the magnetic field and the equipartition arguments for five Galactic SNRs, based on the results empirically obtained from an X-ray data,  can be found in the work of Bamba et al. (2005). The predecessor of this paper is the work of Vukoti{\'c} et al. (2006).

\subsection{Magnetic Field Calculation from Radio Synchrotron Luminosities}

The magnetic field is calculated from the following formula for
synchrotron emission of relativistic electrons (Beck \& Krause
2005, hereafter BK):
\begin{eqnarray}
L_{\mathrm{\nu}}&=&4 \pi f V c_{\mathrm{2}}({\gamma}) n_{\mathrm{e,0}} \cdot \nonumber \\
&& \cdot \ E_{\mathrm{0}}^\mathrm{{\gamma}}
{(\nu/2c_\mathrm{1})}^{(1-\gamma)/2}{H_\mathrm{\perp}}^{(\gamma+1)/2}.
\end{eqnarray}
We adjusted the formula from BK to suit our needs. Here, $f$ is
 a fraction of the radio source volume occupied by the
radiative shell. We assumed that $f=0.25$. This is consistent with
SNRs having strong shocks where compression ratio is $4$. However,
  should this not be the case, variation of $f$ will 
still not  have any significant effect on values for $H$,
because of the small value of exponent $(\gamma+1)/2$ in Eq. (2).
Further, the total volume of SNR is designated  by $V$.
Instead of spectral intensity along the radiation ray path
($I_\mathrm{\nu}$) in BK, we used spectral luminosity of the
source, because the  majority of sources in used data samples
are seen almost as point-like sources, having only the flux
density data integrated over the whole source available. According
to BK this may lead to  the overestimation of values 
for $H$. This effect is discussed further in Section 4. The rest
is the same as in BK, $c_{\mathrm{2}}({\gamma})$ (in units
$\mathrm{{erg^{-2}~s^{-1}~G^{-1}}}$) is identical to
$c_{\mathrm{5}}({\gamma})$ in Pacholczyk (1970),
$n_{\mathrm{e,0}}$ is the number density of cosmic ray electrons
per unit energy interval for the normalization energy
$E_{\mathrm{0}}$, $c_\mathrm{1}=3e/(4\pi
m_\mathrm{e}^3c^5)=6.26428\cdot10^{18} {\rm erg^{-2} ~s^{-1}
~G^{-1}}$, $H_\mathrm{\perp}$ is the magnetic field strength in
the sky plane, and finally $\gamma$ represents exponent in the
cosmic ray  power law energy spectrum (see Appendix A in BK).
  Closer inspection
of Eq. (2)  shows that in order to calculate  $H$ from
$L_\mathrm{\nu}$, some assumption regarding the relationship
between $H$ and $n_{\mathrm{e,0}}$ has to be made.

\subsection{Data Samples}

Currently, it seems that there is no better way to determine $H$
by using only data on $L_\mathrm{\nu}$ and spectral index $\alpha$
(${\gamma=2\alpha+1}$) than the equipartition or the
minimum-energy assumption. This method is useful for SNR samples
where all other data  are lacking. However, Galactic SNR data
samples are strongly biased by selection effects.  The farther
the object, the greater its brightness detection limit. The
extragalactic samples suffer from milder selection effects. Their
brightness detection limits (sensitivity lines) do not differ from one SNR to another because
all the SNRs in  the sample are approximately at the same distance. In this study, we
have  relied on the only statistically reliable sample of
SNRs from a nearby starburst galaxy M82 (Huang et al. 1994). The
equations that we used in calculating $H$ are presented in Section
3. Inspection of those equations  shows that any $H-D$
correlation requires the existence of $L_\mathrm{\nu}-D$
correlation. If $L_\mathrm{\nu}-D$ correlation does not exist,
than it makes no sense to extract $H-D$ relation from
$L_\mathrm{\nu}-D$ data. If SNR data samples show no existence or
poor $L_\mathrm{\nu}-D$ correlation there are two possibilities:
SNR luminosity does not evolve with  the diameter, which is
unlikely, or the sample is made of SNRs that evolve in different
environments and is influenced by selection effects. This is
explained in the next paragraph.

In their work, Arbutina et al. (2004) showed that the best
$L_\mathrm{\nu}-D$ correlation exists for SNRs in M82. They also
showed that some correlation exists for Galactic SNRs associated
with large molecular clouds. Arbutina \& Uro{\v s}evi{\' c} (2005)
  imply that  the evolution of
 SNR radio surface brightness depends on the properties of the
interstellar medium, primarily the density. They formed three SNR
data samples from the existing ones (Galactic and extragalactic):
the Galactic SNRs associated with large molecular clouds (GMC),
oxygen-rich  and Balmer-dominated SNRs. The main  intent of
Arbutina \& Uro{\v s}evi{\' c} (2005) was to group SNRs by their
properties, primarily the density of the interstellar medium in
which they evolve  (and also by SN type). They also argued
that M82  sample is the best possible sample that one can
 currently find. All SNRs from M82 are likely to evolve in
 similar environment of dense molecular clouds. Consequently,
they are very luminous and being extragalactic, SNRs exhibit 
milder selection effects. The reliability of M82 sample is also
discussed in Uro\v{s}evi\'{c} et al. (2005). By performing the
Monte Carlo simulation, the authors showed that M82 sample is not
severely affected by sensitivity selection effects, as in the case
 of other extragalactic samples (LMC, SMC, M31, M33).

In this paper we have  applied REC and calculated $\delta$
for M82 sample, as the  best sample for statistical study,
 and additionally we have analysed three samples from
Arbutina \& Uro{\v s}evi{\' c} (2005): GMC, oxygen-rich and Balmer
dominated  SNRs. We did not use the last three samples to
calculate the slope $\delta$ because they are  of poorer
quality. Instead, we used them  for checking the consistency
of obtained $H$ values with the global picture of SNRs evolution
in different environments. Also, through literature search we
found the magnetic field strengths for some SNRs from Table 2  and
 compared them with the values  obtained in this paper.
We searched the catalog of observational data on Galactic SNRs
from Guseinov et al. (2003, 2004a,b)  and papers available on
the Web-based Astrophysical Data Service ({\it
http://adswww.harvard.edu/}).\footnote[3]{ADS is NASA-funded
project which maintains three bibliographic databases containing
more than 4.7 million records.}

In calculation we have used the radio flux density per unit
frequency interval $S_{\mathrm{\nu}}$ and radio spectral index
$\alpha$ data (Table 2). These two properties are related as:
\begin{equation}
S_\nu=\beta\nu^{-\alpha},
\end{equation}
where $\beta$ is the flux scale factor.  The luminosity is
calculated as $L_\mathrm{\nu}=4 \pi d^2 S_\mathrm{\nu}$, where $d$
is the distance to  an SNR. In the case of extragalactic SNRs
we
 assume that $d$ is the same  for all SNRs, and equal to the distance to the host
galaxy.

\subsection{Magnetic Field and Relativistic Particles}

Since our studies are based on the radio synchrotron luminosity of
SNRs, we can not treat magnetic field separately from
relativistic particles. These two properties of an SNR are
strongly coupled and it makes no sense to study them 
separately.

As mentioned before, calculation of $H$ from Equation (2) requires
an assumption about $n_{\mathrm{e,0}}$.  This quantity also
evolves with $D$. In Table 1 and Section 4.2  we present and
discuss various assumptions about $n_{\mathrm{e,0}}(D)$ evolution
and its effect on $H(D)$ evolution ( assuming empirical $L-D$
relation). Some of the $n_{\mathrm{e,0}}$ evolution patterns are
only illustrative and are used  for estimating the effect of
different patterns on $\delta$. The pattern we used in our method
to calculate  $H$ arises from the equipartition of energies 
implying that energy densities stored in the magnetic field and
relativistic particles are approximately equal. The equipartition
is widely used  for $H$ strength estimates, based purely on
the radio data,  in SNRs, galaxies, etc. It gives reasonably
explainable values for $\delta$ and $H$. Taking all of this into
account we based our method on the equipartition of energies.

Revised equipartition calculation (REC) used to calculate $H$ is
presented in detail in the work of BK. According to BK, REC gives
better results than the classical equipartition calculation (CEC)
presented by Pacholczyk (1970).

\subsection{Evolution of Magnetic Field in SNRs}

 In this subsection we present the theoretical values for
$\delta$ that characterize  a particular SNR evolution phase.
These values,  together with the ones obtained  by our
empirical method, are used in Section 4  in the discussion of
the most probable evolution scenarios for SNRs in M82.

If SNRs are young, in early Sedov or free expansion phase, they
expand practically adiabatically, since radiative energy losses
are negligible. Under the adiabatic expansion assumption i.e.
conservation of energy in cosmic rays and magnetic field
($\frac{d}{dt}(W)=0$), and equipartition conditions
($w_\mathrm{CR}=w_\mathrm{H}$), where $W$ is the total energy and
the quantities $w_\mathrm{CR}$ and $w_\mathrm{H}$ are the energy
 densities of cosmic rays and magnetic field respectively, it
follows that $\delta=1.5$. Indeed:
\begin{equation}
\frac{d}{dt}(W)=\frac{d}{dt}(wV) \propto
\frac{d}{dt}(w_\mathrm{H}V) \propto \frac{d}{dt}(H^2D^3),
\end{equation}
\begin{equation}
\frac{d}{dt}(W)=0\Longrightarrow H \propto D^{-3/2},
\end{equation}
 where $w$ is the total energy density.
  In conclusion, SNRs in the free expansion or early Sedov phase  will have $\delta=1.5$ if they are in the equipartition state.

On the other hand, if SNRs are older, in the late Sedov or
radiative phase,  the value may be closer to $\delta=1.25$.
The radiative phase is characterized by significant energy losses
and SNR would  later expand with the velocity $v \propto
D^{-5/2}$ (pressure-driven snowplow). If $n_\mathrm{e,0} \propto
n_\mathrm{p,0} \propto n_\mathrm{H}v$ (Berezhko \& Volk 2004,
hereafter BV), assuming equipartition $H^2 \propto
n_\mathrm{e,0}$, $\delta$ would be 5/4=1.25.  The quantity
$n_\mathrm{p,0}$  is  the number density of cosmic ray
protons per unit energy interval for the normalization energy
$E_{\mathrm{0}}$, and  $n_\mathrm{H}$ is   the hydrogen
number density.

It is a general belief that, during the expansion, SNRs strongly
amplify interstellar magnetic field. Two basic mechanisms of
magnetic field amplification operate in SNRs. The first one is the
Rayleigh-Taylor instability at the contact discontinuity between
the supernova ejecta and ISM swept by SNR  forward shock.
This scenario leads to $1.5\le\delta\le2$ (Fedorenko 1983) and is
 preferred in young SNRs. The second mechanism operates right
behind the shock, where magnetic field is amplified by strongly
excited magnetohydrodynamic waves. This is the probable mechanism
for older remnants.

\section{Analysis and Results}

There are two most commonly used assumptions regarding  the
magnetic field and cosmic ray energy content: 1) the  minimum of
total energy stored in the particles and magnetic field, and 2)
the equipartition between these energies. The minimum energy
assumption gives $4/3$ for the ratio of the energies stored in the
particles and magnetic field,  which is $\sim 1$. These two
assumptions are thereby often treated as  synonymous and both
procedures are  referred to as the equipartition calculation.
There are also two different methods  for obtaining these two
estimates: classical (Pacholczyk 1970) and revised (BK)
equipartition, i.e. minimum-energy calculation.  We will only
present the formulas that we have used in calculating $H$ and
reader is referred to the mentioned papers for a detailed
treatment of the subject.

\subsection{Classical Calculation}

Classical formulas are:
\begin{eqnarray}
H^\mathrm{min} &=& 4.5^{2/7} {(1 + k)}^{2/7} \cdot \nonumber \\
&&\cdot \ {c_\mathrm{12}}^{2/7} f^{-2/7}
\left({D}/{2}\right)^{-6/7} L^{2/7},
\end{eqnarray}
\begin{eqnarray}
H^\mathrm{eqp} &=& 6^{2/7} {(1 + k)}^{2/7} \cdot \nonumber \\
&&\cdot \ {c_\mathrm{12}}^{2/7} f^{-2/7}
\left({D}/{2}\right)^{-6/7} L^{2/7}.
\end{eqnarray}
In these expressions we have introduced the following quantities:
$k$ is the ratio of the energies of the heavy  relativistic
particles and relativistic electrons, $c_\mathrm{12}$ and
$c_\mathrm{13}$ are functions which are weakly dependent on
$\alpha$ and are tabulated by Pacholczyk (1970). The radio
luminosity $L$ integrated between radio synchrotron spectrum
cutoff frequencies $\nu_\mathrm{1}$ and $\nu_\mathrm{2}$ is
calculated as:
\begin{equation}
L=4 \pi d^2 \int_{\nu_\mathrm{1}=10^7
~\mathrm{Hz}}^{\nu_\mathrm{2}=10^{11}
~\mathrm{Hz}}S_\mathrm{\nu}~d\nu.
\end{equation}
Using Equation (3) we can eliminate $\beta$ and obtain $L$. We
used $k=40$  which should be adequate for strong shocks in SNRs.
Being luminous synchrotron emitters and having small linear
diameters, SNRs from M82 are likely to be young and have strong
shocks, but their true nature is still a subject  to debate.
We obtained $\delta$ from Equations (6) and (7) by replacing $L$
with $L_\nu-D$ relation from Arbutina et al. (2004). Replacing $L$
with $L_\nu$ does not have any noticeable effect on $\delta$. We
also assumed that $H$ depends on $D$ only  trough $L$ or
$L_\nu$.  Therefore,
\begin{equation}
H\propto{\left(D^{-3}L_\nu\right)}^{2/7}\propto{\left(D^{-4.4}\right)}^{2/7},
\end{equation}
if $L(D)\propto D^{-1.4}$ (Arbutina et al. 2004). This gives
$\delta=1.26$. To prove the assumptions in Equation (9), we have
calculated $L$ from equations (3) and (8), and $H$ from Equation
(7). Then we fit linear regression in  $\log H-\log D$ plane to
obtain $\delta=1.26\pm0.08$. This shows that
$c_\mathrm{12}(\alpha)$ does not change with $D$, so it does not
 affect $\delta$,  which is why we can calculate
$\delta$ directly from  the slope $s$ of the $L_\nu\propto D^{-s}$ relation,
\begin{equation}
\delta=(3+s)\frac{2}{7},
\end{equation}
as in Eq. (9).
 Equations (6) and (7) differ to a constant,
giving exactly the same $\delta$. In the sequel we  do not
show results for minimum energy estimates of $H$.

\subsection{Revised Formulas}

The main revision  of the classical formulas is in using $K$
instead of $k$. The quantities $K$ and $k$ stand for the ratios of
proton to electron number densities and energy densities,
respectively. In the CEC, integration of the radiation energy
spectrum between fixed frequency limits is performed.  As
opposed to this in REC the integration is performed over the
energy spectrum of relativistic particles. This gives more
accurate results (see BK).

The revised formulas are:
\begin{eqnarray}
H_{\mathrm{rev}}^{\mathrm{min}} &=& \Big[ 4 \pi K A(\gamma , L_\mathrm{\nu}, \nu ,V ,f, i)\cdot  \nonumber\\
&& \cdot \  C(\gamma , E_\mathrm{2} )( \alpha +1)\Big] ^{1/(
\alpha +3)},
\end{eqnarray}
\begin{eqnarray}
H_\mathrm{rev}^\mathrm{eqp} &=& \Big[ 8 \pi K A(\gamma ,
L_\mathrm{\nu},
\nu , V ,f , i) \cdot  \nonumber\\
&& \cdot \ C( \gamma , E_\mathrm{2} )\Big] ^{1/( \alpha +3)},
\end{eqnarray}
where
\begin{eqnarray}
&C( \gamma , E_\mathrm{2} )= E_\mathrm{0}^2 \cdot \bigg\{
\frac{1}{2} \Big( \frac{E_\mathrm{0}}{E_\mathrm{p}}\Big)
^{\gamma-2} +  & \nonumber\\
  & + \frac{1}{2-\gamma} \bigg[ \Big( \frac{E_\mathrm{0}}{E_\mathrm{2}}
\Big) ^{\gamma-2} -  \Big( \frac{E_\mathrm{0}}{E_\mathrm{p}} \Big)
^{\gamma-2} \bigg] \bigg\} \ \ \  \mathrm{for}\ \gamma \neq 2,\ &
\end{eqnarray}
\begin{equation}
C( \gamma , E_\mathrm{2}
)=E_\mathrm{0}^2\left[\frac{1}{2}+\ln\frac{E_\mathrm{2}}{E_\mathrm{p}}\right]
\ \   \mathrm{for}\ \gamma=2,
\end{equation}
and
\begin{equation}
A(\gamma , L_\mathrm{\nu}, \nu , V, f,
i)=\frac{L_\mathrm{\nu}{(\nu/2c_\mathrm{1})}^{(\gamma-1)/2}}{4 \pi
c_\mathrm{2}(\gamma){E_\mathrm{0}}^\gamma f V c_\mathrm{4}(i)}.
\end{equation}
In  the above equations the following quantities appear: $K$
is the ratio of proton-to-electron number densities per particle
energy interval for the normalization energy $E_\mathrm{0}$,
$E_\mathrm{2}$ presents the high-energy limit for the spectrum of
cosmic ray particles. The spectral break at low energies for
protons is designated  as $E_\mathrm{p}=938.28 \,\mathrm{MeV}
=1.5033\cdot10^{-3} \,\mathrm{erg}$ and finally $c_\mathrm{4}(i)$
is used to replace the projected field component
$H_\mathrm{\perp}$ with the total field $H$ (see Appendix A in
BK), with $i$ being the projection angle.

Equations (11) and (12) were originally taken from BK,  with
a few adjustments. To make equations hold for $\gamma\le2$ we used
$E_\mathrm{2}=3\times~10^{15}~\mathrm{eV}$ (Vink 2004). Instead of
$K+1$ factor we used only $K$ which is justified for
proton-dominated plasma, and since the original formulas do not
include the effect of possible  synchrotron losses that affect the
electron power law energy spectrum. Using $K$ instead of $K+1$
 may provide an even better approximation when taking into
account synchrotron losses.  To put it simple,  it is as
if there were almost no electrons  in cosmic rays, and only
protons  remained. This  can be justified  by the
fact that the protons are far more energetic than electrons and
show less synchrotron losses.  Such assumption does not 
have any significant effect on the values for $H$ because of the
1/($\alpha$+3) exponent in Equations (11) and (12). In this case,
Equation (9) transforms into
\begin{equation}
H\propto{\left(D^{-3}L_\nu\right)}^{1/(\alpha+3)}\propto{\left(D^{-4.4}\right)}^{1/(\alpha+3)},
\end{equation}
and Equation (10) becomes
\begin{equation}
\delta=(3+s)\frac{1}{\overline{\alpha}+3}.
\end{equation}
In Eq. (16) we applied the $L_\nu-D$ correlation, to obtain
$\delta=1.22$, while fitting gives $\delta=1.19\pm0.08$. For
$\alpha$ we used an average spectral index of the whole sample
($\overline{\alpha}=0.6$). The value for $\delta$ from Equation
(17), and  the one obtained by fitting calculated values for
$H$ using Equation (12), are almost  identical. The
difference is  negligible and  we could have smoothly,
as in CEC, calculate $\delta$ from the slope of $L_\mathrm{\nu}-D$
relation.

In calculating $H$, we assumed that the magnetic field in radiative shell of
 SNR is completely turbulent and has an isotropic angle
distribution in three dimensions, giving
$c_\mathrm{4}={(2/3)}^{(\gamma+1)/4}$ (Appendix A in BK). This is
the best assumption  to be made when the majority of SNRs are
point-like sources,  i.e. without maps for $H$. We also used
$K={(E_\mathrm{p} / E_\mathrm{e})}^{(\gamma-1)/2}$ (Appendix A in
BK), where
$E_\mathrm{e}=511~\mathrm{keV}=8.187~10^{-7}~\mathrm{erg}$
designates the spectral break at low energies for electrons. The
data for 21 SNRs from M82  from the work of Uro{\v s}evi{\'
c} et al. (2005), and the obtained values for $H$, are  shown
in Table 2. As it can be seen, the magnetic field strengths are up
to 10 mG. Using $L_\nu/(4 \pi f V)$ in our formulas instead of
$I_\nu/l$ (BK) could lead to an  overestimation of the
average field. Nevertheless, if the magnetic field is
significantly overestimated it should not have  a significant
effect on  the value for $\delta$. There is also a
possibility that M82 remnants are pulsar driven wind nebulae
(PWNe). Unlike shell type SNRs, PWNe have different mechanisms
that maintain magnetic fields. Magnetic field strengths in PWNe
are comparable with the ones we obtained from REC for M82 SNRs.
This possibility is investigated further in Section 3.4.

\subsection{Direct Derivation}

It is possible to derive $\delta$ directly from Eq. (2)  if there
is an additional assumption concerning  the evolution of
$n_\mathrm{e,0}$ with $D$. We consider models used by Shklovsky
(1960), and the assumption of conservation of cosmic ray energy
i.e. adiabatic expansion (e.g. BV). Respectively,  these are
\begin{equation}
n_\mathrm{e,0} \propto D^{-(2\alpha+3)}
\end{equation}
and
\begin{equation}
n_\mathrm{e,0} \propto D^{-3}.
\end{equation}
Equation (2) together with $L_\nu-D$ relation gives
\begin{equation}
H \propto
{\left(\frac{D^{-4.4}}{n_\mathrm{e,0}}\right)}^{1/(\alpha+1)}.
\end{equation}
For an average spectral index $\alpha = 0.6$ the results are
presented in Table 1. Here, we found fitting  unnecessary because
we already saw in Sections 3.1 and 3.2 that the rest of the
quantities from Equation 2 do not change with $D$, at least 
not in a way to affects $\delta$.  By using direct method we
can only get values for $H$ scaled to a constant because of
proportionality  of Equations (18) and (19).

\begin{Code}
  \begin{table}
    \caption{RESULTS FOR $\delta$ \label{label}}
  \begin{center}
      \begin{tabular}{cc}\hline\hline
\multicolumn{2}{c}{\textbf{Direct}}\\
\hline
Shklovsky (1960)  ($n_\mathrm{e,0} \propto D^{-(2\alpha+3)}$)&0.125\\
Berezhko \& Volk (2004) ($n_\mathrm{e,0} \propto D^{-3}$)&0.875\\
\hline
\multicolumn{2}{c}{\textbf{Classical}}\\
\hline
equipartition&$1.26$\\
\hline
\multicolumn{2}{c}{\textbf{Revised}}\\
\hline
equipartition&1.22\\
 \hline\hline
\end{tabular}
    \end{center}
    \end{table}
\end{Code}

\subsection{Calculated and Literature-found $H$ Values for GMC, Oxygen-rich and Balmer-dominated SNRs}

With a view to checking values obtained for $H$ we performed
the same REC on the SNRs associated with large molecular clouds,
oxygen-rich and Balmer dominated SNRs. According to Arbutina \&
Uro{\v s}evi{\' c} (2005), these SNRs form parallel tracks in
radio surface brightness to diameter plane. If environmental
density is higher  we expect the SNR to be brighter.  The
implication is that SNRs with the same $D$ should have different
luminosities if environmental densities are different. According
to Equation (12), SNRs that evolve in  a more dense
environment should also have stronger $H$ than SNRs with the same
diameter that evolve in a less dense environment. The data used
and the obtained CEC and REC results for all groups of SNRs are
presented in Table 2.

Figure 1 presents a plot of all REC values from Table 2. It shows
that SNRs in  a more dense environment (M82, GMC,
oxygen-rich)  appear to form a track in $H-D$ plane, while
Balmer-dominated SNRs form another track that lies beneath the
first one. Due to  the dispersion and incompleteness of data
samples, any statistical study of the tracks should be avoided,
for now.  We can, however, make some qualitative conclusions.
In Figure 1 we can see that REC does not change $L_\nu-D$
 the evolution pattern. This is very convenient for the
estimate of  the reliability of $H$ in M82 SNRs. From Figure
1 it is clear that $H$ values for SNRs in M82 seem consistent with
the values for GMC and oxygen-rich remnants. They all evolve in
 a dense environment and accordingly may have  a similar
$H-D$ evolution pattern. Their $H$ values, according to Arbutina
\& Uro{\v s}evi{\' c} (2005), are different in comparison to the
values for Balmer-dominated SNRs. This is because Balmer-dominated
SNRs {are likely to evolve} in  a low density environment. In
the group that consists of Balmer-dominated, oxygen-rich and GMC
SNRs, used in this work, we  did not include PWNe, because
REC is made for shell type SNRs. Accordingly, to avoid possible PWNe, we did not include SNRs with $\alpha\le 0.4$, which is the characteristic of PWNe (Gaensler \& Slane 2006). From Figure 1 we can
see that the most of SNRs in M82 are, probably, not PWNe because
they fit the evolution pattern for SNRs in dense environments. In
addition, the higher spectral indices of the M82 SNRs (average
$\alpha\approx0.6$; see Table 2) are not characteristic for PWNe.
However, the possibility that  at least some of these objects
are PWNe should not be easily  put aside. For now, we can
only wait for  the observational instruments to advance, and for a
possible detection of pulsars in M82.

Table 2 also shows the best available literature-found
values for $H$ inferred from other methods, for Galactic and LMC
SNRs. The agreement of these  values with the values obtained
from REC  is another way to show the reliability of $H$
estimates for SNRs in M82. This is one of the subjects 
discussed in Section 4.

\begin{Code}
  \begin{figure}
    \begin{center}
      \includegraphics[height=7.8cm, width=8.3cm]{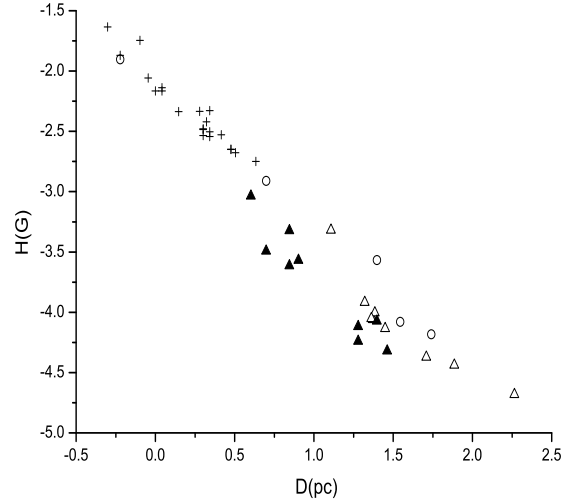}
    \end{center}
  \caption{The revised equipartition data in  $\log
H-\log D$ plane. The SNRs are presented by: crosses (M82), open circles (oxygen-rich), open triangles (Galactic SNRs associated with large molecular clouds), filled triangles (Balmer-dominated SNRs). \label{label} }
  \end{figure}
\end{Code}

\section{Discussion}

\subsection{Values Obtained  for $H$}

Both, classical and revised equipartition calculation
contain various uncertainties and assumptions and  as such,
are of limited applicability (BK). Nevertheless, by performing
CEC and REC
 we  arrived at the conclusion that all of the imperfections do not
have  a noticeable effect on $\delta$, but could have a
significant  impact on the values for $H$. Inspection of
Table 2 shows that obtained $H$ values are higher than those 
found in literature. Such overestimates are probably due to
replacement of $I_\mathrm{\nu}$ with $L_\mathrm{\nu}$ (BK).  The
assumptions regarding $f$ and $K$ in REC equations are not of
great  importance because of the small $1/( \alpha +3)$
exponent. Due to  the $L_\mathrm{\nu} \rightarrow
I_\mathrm{\nu}$ replacement, the amount of overestimate is
strongly affected by SNR morphology and consistently shows 
considerable variations from one SNR to another (Table 2). The
morphology variations should not depend on diameter,  which
means that overestimates of $H$ are mainly arising from morphology
related factors, and they should only make data scattering in
$H-D$ plane  without affecting $\delta$. Table 2  also
shows that an average overestimate by a factor of 2 can be
adopted. Coupled with explanation of Figure 1 (Section 3.4), this
shows that $H$ values for SNRs in M82 are estimated reliably to an
order of magnitude. This means that M82  does contain SNRs
with magnetic fields of up to $10^{-2}$ G. However, this should be
taken with reserve because of the possibility that some SNRs in
M82 are  perhaps PWNe.

REC used in this paper  thus gives reliable estimates
accurate to an order of magnitude. This is of  small
significance in studies of nearby, well resolved SNRs with data
from all parts of electromagnetic spectrum,  but may be of
great applicability in statistical and empirical studies of SNRs
 residing in other galaxies,  that are unresolved and
often have only radio data available.  As  already mentioned,
Galactic SNR samples are strongly influenced by selection effects
and can not be used in  the statistical and empirical studies
of SNRs evolution properties. For now,  the only SNR samples
that can be used for reliable statistical and empirical studies
reside in other galaxies.  With these samples, the obtained
values for $H$ will be probably overestimated by a factor of 2,
but accurate to an order of magnitude,  as in this paper. In
the next section we discuss  the results on the magnetic
field evolution obtained when our method is applied to SNRs in
M82. This should illustrate  how the method can be used for
getting closer insight into the SNRs evolution properties, i.e.
SNRs evolution phases,  and how it can be used to check the
validity of the equipartition assumption.

\subsection{Magnetic Field Evolution of SNRs in M82}

If the sample is statistically reliable, the obtained $H$  may
be overestimated,  but chosen REC parameter values should
not have  a significant effect on $\delta$. The difference
between $\delta$ obtained from classical and revised methods is
mainly due to the exponents in equations (7) and (12). These
exponents will be equal for an average spectral index
$\alpha=0.5$. For SNRs in M82 $\bar{\alpha}=0.6$ is used, and
therefore we obtain slightly different slopes in $H-D$ plane. In the work of Berkhuijsen (1986), the author implies that $\alpha$ could depend on the density of the ISM in which the SNRs evolve as: $\alpha=(0.075\pm 0.024)\log n_\mathrm{0}+(0.538\pm 0.012)
$, where $n_\mathrm{0}$ is the density of the ISM. This means, according to Eq. (17), that the lower track in Figure 1, that consist of SNRs evoloving in the small density environments, should have somewhat shallower slope when compared with the track above (large density environments).  However, cosidering Eq. (17) and just mentioned relation, it is clear that for typical values of $\alpha$ and $n_\mathrm{0}$, there will be no significant effect on $\delta$. Consequently, the tracks from Fig. 1 should be considered as parallel. 
Taking all of above into account, we conclude that $\delta$ is
strongly affected by the assumptions regarding $n_\mathrm{e,0}$.
"Directly" obtained values for $\delta$ of 0.125 and 0.875 (Table
1) are only illustrative. Shklovsky's model have a rather
historical meaning, since no additional particle acceleration (by
the shock) during evolution is assumed ( besides the initial
acceleration in supernova explosion).  This leaves us with the
equipartition as our best assumption.

Table 1 shows that the equipartition arguments combined with the
possible $L_\nu-D$ dependence give $\delta\approx1.2$. This value
is slightly lower than theoretical value $\delta=1.5$ obtained
under equipartition and adiabatic approximations (Section 2.5). If
SNRs in M82 are young, in early Sedov or free expansion phase,
 this difference can be explained by  the sensitivity selection
effects related to  the M82 sample. The Monte Carlo
simulations in Uro{\v s}evi{\' c} et al. (2005) show that 
the measured slopes of extragalactic surface brightness to
diameter ($\Sigma_\nu-D$) relations are shallower due to  the
sensitivity selection effects. Therefore, the apparent $\Sigma_\nu-D$ (
and $L_\nu-D$) slope for M82 is lower than the real slope. The lower
$L_\nu-D$ slope gives lower $\delta$. This means that
equipartition arguments for the SNRs in M82 sample may  still
be applicable,  whereas a small difference between  the
theoretical and empirical $\delta$ can be  ascribed to
selection effects.

On the other hand, $\delta=1.2$ might indicate that not all SNRs
from M82 sample are in the equipartition stadium. If, for example,
the larger ones are in the late Sedov phase where magnetic field
remains constant (BV), empirical $\delta$ would be a compromise
between values 0 and 1.5. The evolutionary status of SNRs
  remains a great uncertainty. The SNRs in M82 may be in the
free expansion, as well as in the Sedov's, or even in the
radiative phase. Chevalier \& Fransson (2001) proposed that M82
SNRs may be in the radiative phase  because they evolve in
 a very dense environment. In this case, $\delta$ may be
1.25, close to the empirical value.  As the previous ones,
this scenario too,
 remains uncertain.

\subsection{Interstellar Magnetic Field in M82}

Condon (1992) estimated field strength in M82  to be
$H\approx100~\mathrm{\mu G}$ from classical minimum energy
calculation, considering that central emitting region of M82 is
$30''\times10''$ and probably 0.5 kpc thick. Hargrave (1974)
estimated the central emitting region in M82 to be $50''\times
15''$. Using revised equipartition we estimated the value of
${\approx 190}~\mathrm{\mu G}$ for the average interstellar
magnetic field in the central emitting region of M82 using the
data $S_\mathrm{1.4~GHz}=8.2~ \mathrm{Jy}$ and $\alpha=0.68$ from
Klein et al. (1988). We assumed that $f=1$ and that M82 radiates
mainly from its central region of $\approx500~\mathrm{pc}$ in
diameter. This estimate is  rough and should be taken with
 some reserve. Such ISM magnetic field strength is among the
highest field strengths when compared  to other galaxies.
This,  however, may imply that M82  central region
contains interstellar matter  made of very dense molecular
clouds. This is consistent with the high values of $H$ in M82
SNRs, supporting the possibility that their luminous
synchrotron emission is mainly due to very dense environments and
not due to pulsar driven wind nebulae.

The values of up to 10 mG for $H$ in M82 SNRs however imply that
 the magnetic field is strongly amplified from the average
ISM values of $100-200~\mathrm{\mu G}$.

\section{Conclusions}

In this paper we presented and discussed  a method for 
the determination of the magnetic field evolution pattern in SNRs
only from the radio luminosity data samples. Such samples are the
only available for the statistical and empirical studies of SNR
evolution properties.  The best sample, for now, consists of SNRs
in M82, since these remnants  seem to evolve in similar
environment and share similar properties,  and are not severely
influenced by selection effects.

In order to calculate $H$ from REC we were forced to make some
assumptions. The only significant effect on values for $H$,
regarding the assumptions, comes from replacing $I_\mathrm{\nu}$
in REC formulas from BK with $L_\mathrm{\nu}$, which is done in
order to apply REC on  practically point-like sources. The
other assumptions are  less important because of the small
exponent in REC equations. Obtained under equipartition
assumption, $\delta$ is a direct consequence of $L_\mathrm{\nu}-D$
slope and has reasonable theoretical explanation. All assumptions
do not change the evolutionary picture from $L_\mathrm{\nu}-D$
plane. This means that our empirical estimate of $\delta$ is
likely to be  reliably  determined. When compared with the
more reliable values found in literature, the obtained $H$ values
 appear to be overestimated approximately by a factor of 2.
We conclude that $H$ values for all SNRs, even the ones from M82,
are accurate to an order of magnitude.

To answer whether or not M82 SNRs are in equipartition state we
have compared empirical  $\delta$ obtained by our method with the
theoretical values. The empirically obtained $\delta$ from
$L_\mathrm{\nu}-D$ correlation under the equipartition assumption
is probably theoretically explainable by the following two
scenarios:

(i) The slight difference between  the theoretically derived
$H-D$ slope ($\delta=1.5$) under the adiabatic approximation and
the equipartition assumption, and the slope obtained in this paper
using the empirical $L_\mathrm{\nu}-D$ correlation and  REC
($\delta\approx1.2$) can be explained by the sensitivity selection
effects which affected the sample of SNRs in M82. In this way, the
starting assumption concerning the approximative equipartition
between  the energy stored in the relativistic particles and
in the magnetic field, could be justified. Therefore, we can
conclude that SNRs in the M82 sample are probably  close to
the equipartition state.

(ii) Finally, equipartition conditions may not be fulfilled for
all remnants.  If, for instance, they are in different stages
of evolution, $\delta$ may be between 0 and 1.5.

If SNRs are in the adiabatic phase, the most probable explanation
for the lower empirically obtained value for $\delta$  are the
sensitivity selection effects in the M82 sample, perhaps in
combination with slight deviation from equipartition, but the
problem  is the unresolved evolutionary status of M82 SNRs.
Additional observations of SNRs in nearby starburst galaxies are
needed for  any firmer conclusions to be made.

\begin{Code}
  \begin{table*}
    \caption{SNRs DATA\tabnotemark{a} AND RESULTS \label{label}}
  \begin{center}
\small
      \begin{tabular}{@{\extracolsep{-0.7mm}}p{2.4cm}p{1.7cm}p{1.0cm}cccccccc@{}}
\hline\hline
 Catalog&Other&Type\tablenotemark{1}&$D$&$S_\mathrm{1}$&$\alpha$&Distance&$H^\mathrm{eqp}$&$H_\mathrm{rev}^\mathrm{eqp}$&$H_\mathrm{l}$\\
name&name&&&\small{flux density}&&&\small{class.}&\small{rev.}&\small{literature}\\
&&&&\small{at 1 GHz}&&&\small{equip.}&\small{equip.}&\\
&&&(pc)&(mJy)&&(kpc)&(G)&(G)&(G)&\\
\hline
M82 39.1+57.4&\nodata&MC&0.9&8.28&0.50&$3.9\times10^3$&6.03E-03&8.76E-03&\nodata\\
M82 39.4+56.1&\nodata&MC&3.23&4.25&0.58&$3.9\times10^3$&1.68E-03&2.10E-03&\nodata\\
M82 39.6+53.4&\nodata&MC&2.65&2.68&0.45&$3.9\times10^3$&1.74E-03&2.96E-03&\nodata\\
M82 40.6+56.1&\nodata&MC&3.02&4.97&0.72&$3.9\times10^3$&1.94E-03&2.24E-03&\nodata\\
M82 40.7+55.1&\nodata&MC&1.93&15.56&0.58&$3.9\times10^3$&3.78E-03&4.64E-03&\nodata\\
M82 41.3+59.6&\nodata&MC&1.02&6.19&0.52&$3.9\times10^3$&4.99E-03&6.85E-03&\nodata\\
M82 42.7+55.7&\nodata&MC&4.30&6.10&0.71&$3.9\times10^3$&1.51E-03&1.78E-03&\nodata\\
M82 42.8+61.3&\nodata&MC&1.97&3.58&0.63&$3.9\times10^3$&2.47E-03&2.92E-03&\nodata\\
M82 43.2+58.4&\nodata&MC&1.05&12.61&0.66&$3.9\times10^3$&6.11E-03&6.83E-03&\nodata\\
M82 43.3+59.2&\nodata&MC&0.60&29.54&0.68&$3.9\times10^3$&1.27E-02&1.35E-02&\nodata\\
M82 44.3+59.3&\nodata&MC&1.96&5.46&0.64&$3.9\times10^3$&2.80E-03&3.27E-03&\nodata\\
M82 44.5+58.2&\nodata&MC&2.25&3.55&0.50&$3.9\times10^3$&2.16E-03&3.13E-03&\nodata\\
M82 45.2+61.3&\nodata&MC&1.12&19.54&0.67&$3.9\times10^3$&6.58E-03&7.28E-03&\nodata\\
M82 45.3+65.2&\nodata&MC&2.05&5.80&0.82&$3.9\times10^3$&2.96E-03&3.32E-03&\nodata\\
M82 45.4+67.4&\nodata&MC&2.23&5.01&0.67&$3.9\times10^3$&2.47E-03&2.86E-03&\nodata\\
M82 45.8+65.3&\nodata&MC&2.13&3.74&0.46&$3.9\times10^3$&2.30E-03&3.79E-03&\nodata\\
M82 45.9+63.9&\nodata&MC&2.22&4.25&0.41&$3.9\times10^3$&2.32E-03&4.70E-03&\nodata\\
M82 46.5+63.9&\nodata&MC&1.39&6.93&0.74&$3.9\times10^3$&4.18E-03&4.60E-03&\nodata\\
M82 46.7+67.0&\nodata&MC&2.95&4.39&0.76&$3.9\times10^3$&1.94E-03&2.25E-03&\nodata\\
M82 41.9+58.0&\nodata&MC&0.52&154.96&0.75&$3.9\times10^3$&2.38E-02&2.32E-02&\nodata\\
M82 44.0+59.6&\nodata&MC&0.79&54.89&0.48&$3.9\times10^3$&1.16E-02&1.80E-02&\nodata\\\hline
G 111.7-2.1&Cas A&O&4.9&$2720\times10^3$&0.77&3.4&1.02E-03&1.23E-03&5.5E-04\tabnotemark{b}\\
G 260.4-3.4&Pup A&O&35.2&$130\times10^3$&0.5&2.2&5.73E-05&8.32E-05&\nodata\\
LMC 0525-69.6&N132 D&O&25&5800&0.7&55&2.07E-04&2.71E-04& $<$ 4E-05\tabnotemark{c}\\
SMC 0103-72.6&\nodata&O&55&250&0.5&65&4.53E-05&6.58E-05&\nodata\\
NGC 4449&\nodata&O&0.6&20&0.75&4200&1.22E-02&1.25E-02&\nodata\\\hline
G 42.8+0.6&\nodata&MC&76.8&$3\times10^3$&0.5&11&2.51E-05&3.64E-05&\nodata\\
G 78.2+2.1&$\gamma$ Cygni&MC&20.9&$340\times10^3$&0.5&1.2&8.34E-05&1.21E-04&\nodata\\
G 84.2-0.8&\nodata&MC&23.6&$11\times10^3$&0.5&4.5&6.00E-05&8.71E-05&\nodata\\
G 89.0+4.7&HB 21&MC&24.2&$220\times10^3$&0.4&0.8&5.20E-05&9.87E-05&\nodata\\
G 132.7+1.3&HB 3&MC&51.2&$45\times10^3$&0.6&2.2&3.10E-05&4.24E-05&\nodata\\
G 166.2+2.5&OA 184&MC&183.8&$11\times10^3$&0.57&8&1.43E-05&2.08E-05&\nodata\\
G 309.8+0.0&\nodata&MC&23&$17\times10^3$&0.5&3.6&6.11E-05&8.88E-05&\nodata\\
G 315.4-2.3&MSH 14-63&MC&28.1&$49\times10^3$&0.6&2.3&5.45E-05&7.34E-05&\nodata\\
G 349.7+0.2&\nodata&MC&8.7&$20\times10^3$&0.5&14.8&3.3E-04&4.80E-04&3.5E-04\tabnotemark{d}\\\hline
G 4.5+6.8&Kepler&B&2.4&$19\times10^3$&0.64&2.9&3.95E-04&4.97E-04&2.15E-04\tabnotemark{b}\\
G 120.1+1.4&Tycho&B&5&$56\times10^3$&0.61&2.3&2.49E-04&3.21E-04&3E-04\tabnotemark{b}\\
G 327.6+14.6&SN 1006&B&19&$19\times10^3$&0.6&2.2&5.67E-05&7.62E-05&1.6E-04\tabnotemark{b}\\
LMC 0505-67.9&DEM L71&B&19&9&0.5&55&3.96E-05&5.75E-05&\nodata\\
LMC 0509-68.7&N103 B&B&7&1100&0.6&55&3.72E-04&4.75E-04&\nodata\\
LMC 0509-67.5&\nodata&B&7&70&0.5&55&1.68E-04&2.43E-04&\nodata\\
LMC 0519-69.0&\nodata&B&8&150&0.5&55&1.86E-04&2.70E-04&\nodata\\
LMC 0548-70.4&\nodata&B&25&100&0.6&55&6.29E-05&8.44E-05&\nodata\\
SMC 0104-72.3&\nodata&B&29&12&0.5&65&3.29E-05&4.78E-05&\nodata\\\hline\hline

\end{tabular}
    \end{center}
\vspace{3mm} \footnotesize{ Notes: $^\mathrm{a}$M82 data are taken
from Table A.1 in Uro{\v s}evi{\' c} et al. (2005) with
$S_\mathrm{1}$ being scaled from 1.4 to 1 GHz. The rest of the
used data are same as in papers of  Arbutina et al. (2004) and
Arbutina \& Uro{\v s}evi{\' c} (2005), with data of Galactic MC
SNRs being updated for distances from Green (2004);
 $^\mathrm{b}$V{\" o}lk et al. (2005); $^\mathrm{c}$Dickel \& Milne (1995); $^\mathrm{d}$Brogan et al. (2000); $^\mathrm{1}$MC -- Associated with giant molecular clouds, O -- Oxygen-rich, B -- Balmer-dominated.
}
\end{table*}
\end{Code}

\acknowledgments

\noindent {\it Acknowledgments.} The authors thank Dragana Momi{\'c} and Ivanka Mutavd{\v z}i{\'c} for careful reading and correction of the manuscript, and an anonymous referee for comments and suggestions. The authors would also like to thank
Prof. Rainer Beck for useful comments on the manuscript. This paper
is a part of the project "Gaseous and Stellar Components of
Galaxies: Interaction and Evolution" (No. 146012) supported by the
Ministry of Science and Environmental Protection of Serbia.

\end{document}